\documentclass{article}
\usepackage{graphicx}
\usepackage{lineno}
\usepackage{placeins}
\usepackage{amsmath}
\usepackage[a4paper, total={6in, 8.25in}]{geometry}

\newcommand\pubdate{\today}

\def\Title#1{\begin{center} {\Large #1 } \end{center}}
\def\Author#1{\begin{center}{ \sc #1} \end{center}}
\def\Address#1{\begin{center}{ \it #1} \end{center}}

\newcommand\pubblock{\rightline{\begin{tabular}{l}  \\ 
         \pubdate  \end{tabular}}}
\newenvironment{Abstract}{\begin{quotation}  }{\end{quotation}}
\newenvironment{Presented}{\begin{quotation} \begin{center} 
             PRESENTED AT\end{center}\bigskip 
      \begin{center}\begin{large}}{\end{large}\end{center} \end{quotation}}

\begin{document}
\begin{titlepage}
 \pubblock
\vfill
\Title{Measurements of the Higgs boson coupling properties to fermions with the ATLAS detector}
\vfill
\Author{ Khuram Tariq\footnotemark[1],\\on behalf of the ATLAS collaboration\footnotemark[2] }
\Address{Institute of High Energy Physics, Chinese Academy of Sciences, Beijing, China}
\vfill

\begin{Abstract}
Testing the Yukawa couplings of the Higgs boson to quarks and leptons is important to understand the origin of fermion masses. These proceedings will review several measurements of Higgs boson decays to two bottom quarks or two tau leptons, searches for Higgs boson decays to two charm quarks or two muons, as well as direct constraints on the charm-Yukawa coupling. The production of Higgs boson in association with top quarks will also be discussed. These analyses are based on 139 fb$^{-1}$ of Run-2 data from proton-proton collisions collected by the ATLAS experiment at the Large Hadron Collider (LHC) with a center-of-mass energy of 13 TeV.
\end{Abstract}
\vfill

\begin{Presented}
DIS2023: XXX International Workshop on Deep-Inelastic Scattering and
Related Subjects, \\
Michigan State University, USA, 27-31 March 2023 \\
     \includegraphics[width=9cm]{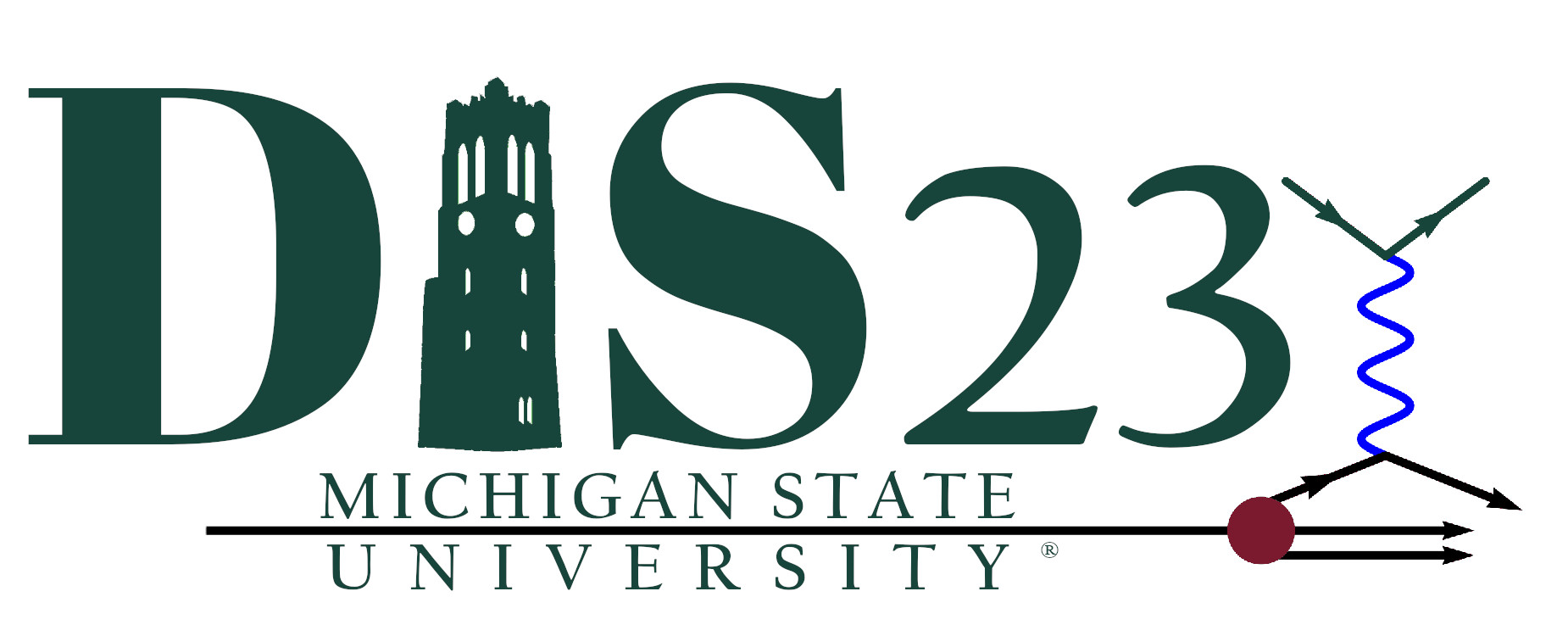}
\end{Presented}
\vfill
\footnotetext[1]{Email: khuram.tariq@cern.ch}
\footnotetext[2]{\copyright Copyright 2023 CERN for the benefit of the ATLAS Collaboration. CC-BY-4.0 license.}
\end{titlepage}

\section{Introduction}
Higgs couplings with a massive elementary particle is determined by the particle mass and type. Experimental measurements of the Higgs boson couplings to fermions provide important tests of the Standard Model and stringent constraints on theories beyond it. Recent studies, performed at ATLAS experiment~\cite{ATLAS} during the Run 2 of the Large Hadron Collider (LHC) have successfully measured the interactions with the third-generation fermions (bottom and top quarks, and tau leptons), while emerging indications of interactions with the second-generation particles (muons and charm quarks) are to be observed. These measurements contribute to our understanding of fundamental physics and offer potential insights into new physics phenomena. 

Most of these differential measurements are based on the Simplified Template Cross Section (STXS) framework~\cite{STXS} which defines specific kinematic regions per production process to be measured. There are different methods to interpret the STXS measurement results. One of the most promising tools in this field is the Effective Field Theory (EFT) framework~\cite{EFT}, which provides systematic approaches to describe and analyze physics beyond observable energy scales by incorporating effective theories that capture relevant degrees of freedom and interactions.

\section{Higgs Boson Production and Decay}
The Higgs boson is predominantly produced through the gluon-gluon fusion (ggF) at LHC, accounting for 87\% of the production, followed by the vector boson fusion (VBF) at 7\%, the associated weak gauge boson production $(VH)$ at 4\%, the associated top-antitop production $(t\bar{t}H)$ at 1\%, and other production modes with small cross sections~\cite{corss-section}. The measured mass of the Higgs boson is 125.09 GeV~\cite{higgs}, and in the Standard Model (SM), it predominantly decays to $b\bar{b}$ pairs with a branching ratio (BR) of approximately 58.1\%~\cite{corss-section}. However, for certain measurements, higher sensitivity is achieved by focusing on final states with higher purity and resolution, such as $H\to\tau\tau$ decay with a BR of around 6.26\%, $H\to\gamma\gamma$ decay with a BR of approximately 0.227\%, and $H\to ZZ^*\to$ $\ell \ell$ $\ell’ \ell’$ $(\ell/\ell’ = e, \mu)$ decay with a branching ratio of only about 0.0125\%~\cite{corss-section}.

\section{Measurements of $H\to b\bar{b}$ via associated $VH$ production}
The observation of the decay of the Higgs boson into $b$-quark pairs provided direct evidence for the Yukawa coupling of the Higgs boson to down-type quarks. The ATLAS Collaboration has published two new measurements of $H\to b\bar{b}$ in $V(=W^{\pm}/Z^0)H$ production. The first measurement, known as the resolved $VH(b\bar{b})$~\cite{hbb1}, reconstructs leptonic $W/Z$-boson decays and requires exactly two $b$-tagged anti-$k_t$ R=0.4 jets to identify the Higgs boson candidate. The second measurement, referred to as the boosted $VH(b\bar{b})$~\cite{hbb2}, adopts the same reconstruction strategy for the leptonically decaying $W/Z$-boson but reconstructs the Higgs boson using an anti-$k_t$ R=1.0 jet (large-R) along with two matched $b$-tagged track-jets. Track-jets are built with the anti-$k_t$ algorithm with a variable radius (VR) $p_\text{T}$-dependent parameter, from tracks reconstructed in the inner detector. The resolved $VH(b\bar{b})$ measurement focuses on a phase space region where the $p_\text{T}^H$ (transverse momentum of the Higgs candidate) is above 60 GeV, while the boosted $VH(b\bar{b})$ measurement targets $p_\text{T} \ge$ 250 GeV.

\begin{figure}[t]
	\centering
	\includegraphics[width=7cm]{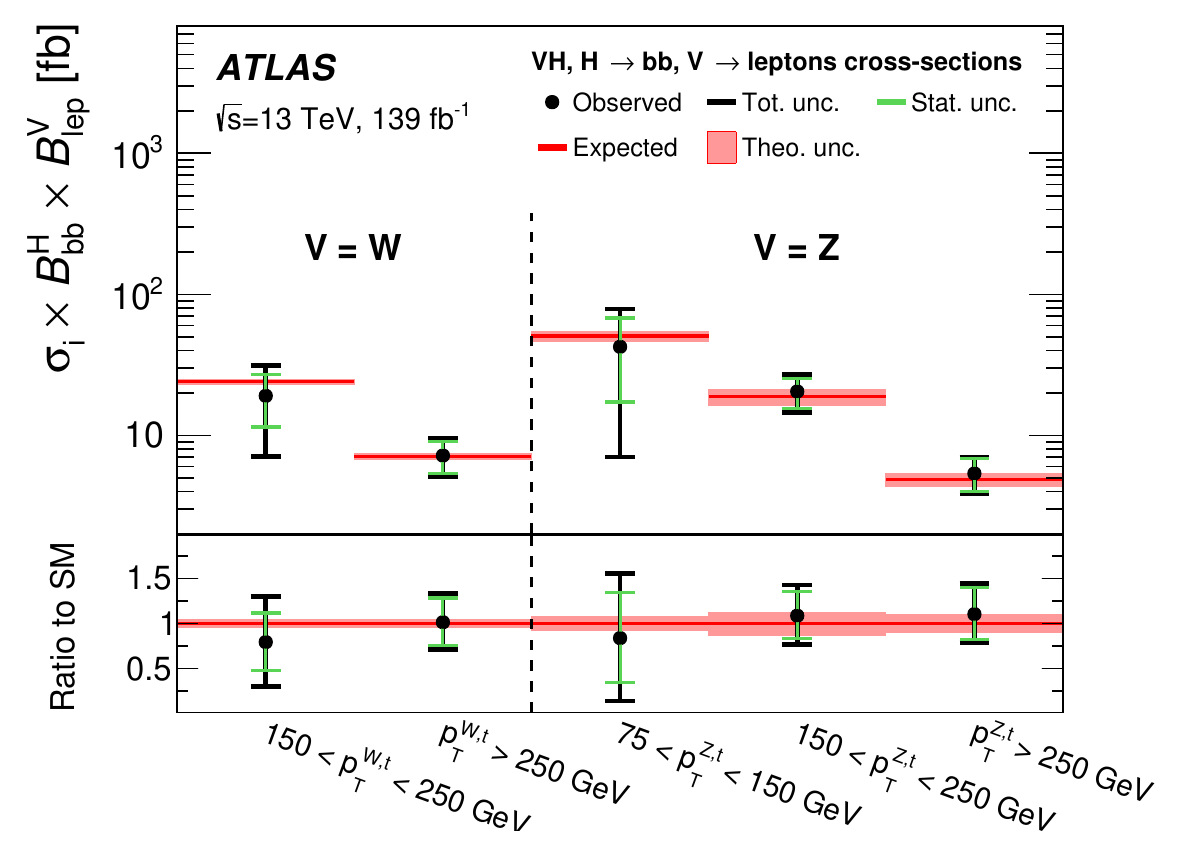}
	\includegraphics[width=7cm]{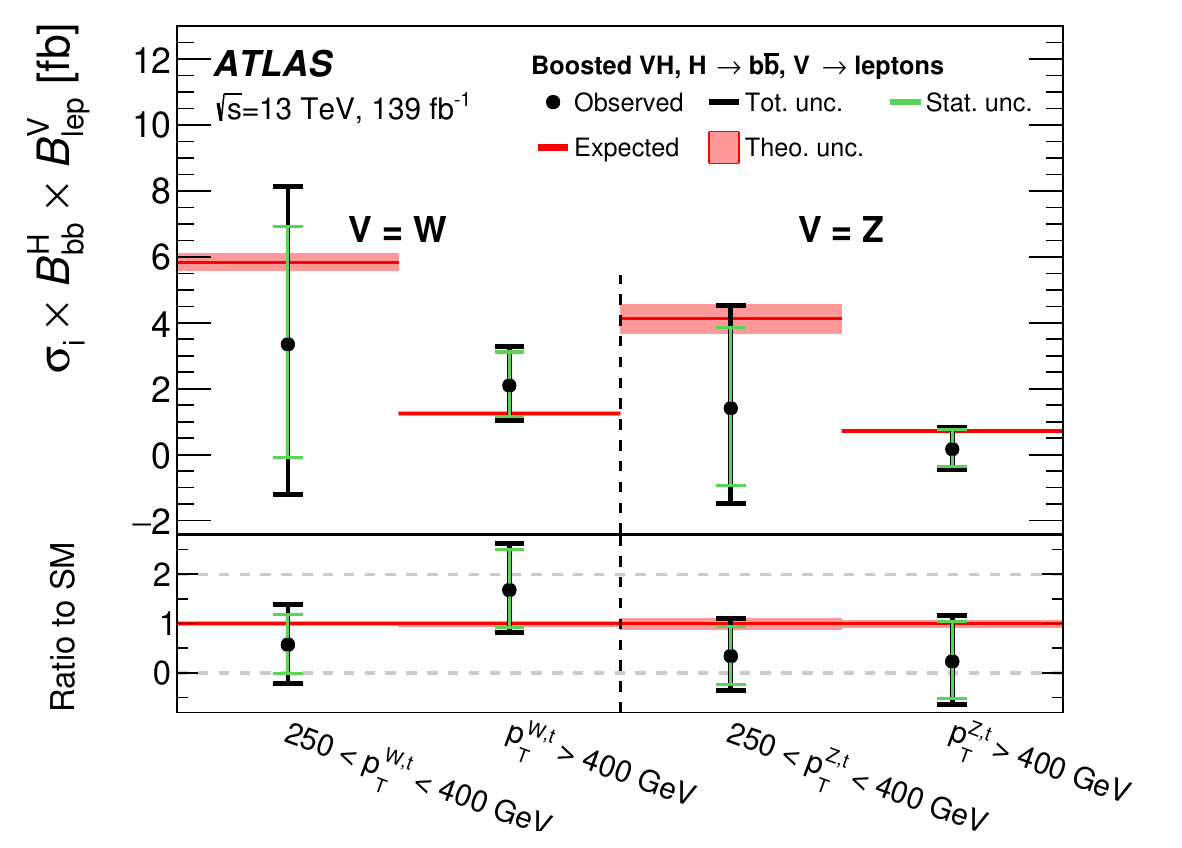}
	\includegraphics[width=7cm]{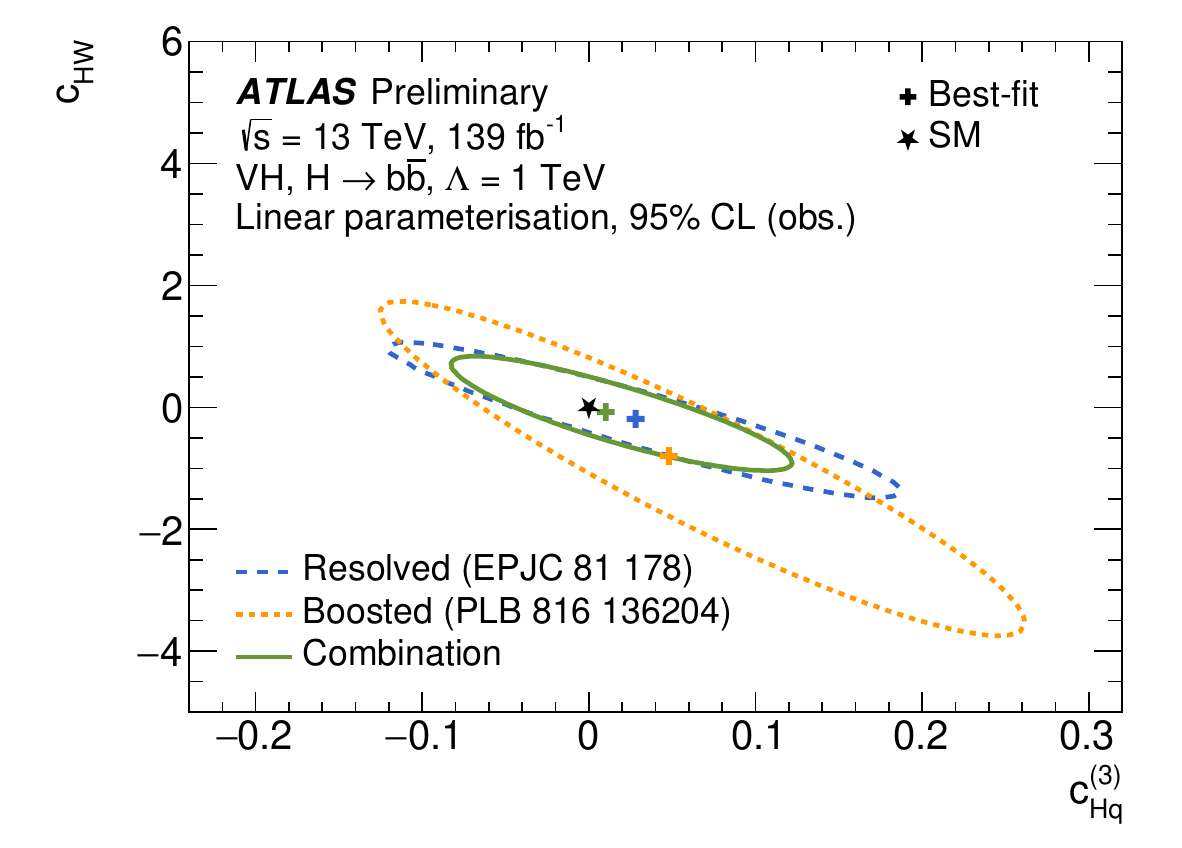}
	\caption{Measured $VH(b\bar{b})$ simpliﬁed template cross sections times the $H\to b\bar{b}$ and $V$ → leptons branching fractions (top-left: resolved~\cite{hbb1}, top-right: boosted~\cite{hbb2}). The 95\% confidence level (CL) contour for the simultaneous fit of the $c_{Hq}^{(3)}$ and $c_{WH}$ coefficients when considering the linear terms (bottom)~\cite{hbb3}.}
	\label{fig:hbb}
\end{figure}

Both analyses incorporate measurements based on the STXS framework and interpretations within the EFT. The measured $VH$ cross section times branching fraction $\sigma \times B$ in each STXS bin, together with the SM predictions are summarised in Figure~\ref{fig:hbb}  for the resolved (top-left) and boosted (top-right) analysis categories. In the resolved $VH(b\bar{b})$ analysis, the observed value of $\mu^{b\bar{b}}_{VH}$ is $1.02^{+0.12}_{-0.11}(stat.)^{+0.14}_{-0.13}(syst.)$, indicating a rejection of the background-only hypothesis with a significance of 6.7 standard deviations. The analysis also provides separate results for $WH$ and $ZH$ production modes, which exhibit data excesses above the background-only hypothesis with significances of 4.0 and 5.3 standard deviations, respectively. In the boosted $VH(b\bar{b})$ analysis, an observed value of $\mu^{b\bar{b}}_{VH}$ = $0.72^{+0.29}_{-0.28} (stat.)^{+0.26}_{-0.22}(syst.)$ is obtained, indicating a data excess above the background-only hypothesis by 2.1 standard deviations. The combination of both measurements~\cite{hbb3} provide  more complete picture of the $VH$ process. The orthogonality of the analyses is ensured through the implementation of $p_\text{T}$ cut at 400 GeV. This allows for a more detailed and precise measurement of the STXS in a more granular manner. The results obtained from these combined analyses are well understood and found to be compatible with the individual analyses. Figure~\ref{fig:hbb} (bottom) shows two dimensional profile likelihood scans for Wilson coefficients, $c_{Hq}^{(3)}$ vs. $c_{WH}$ using the linear parametrization. Only the interference effect between the SM diagram and diagrams corresponding to beyond SM (BSM)  dimension-six operators in the EFT Langrangian is considered.

\section{Higgs-Top coupling ($t\bar{t}H$)}
\begin{figure}[t]
	\centering
	\includegraphics[width=7cm]{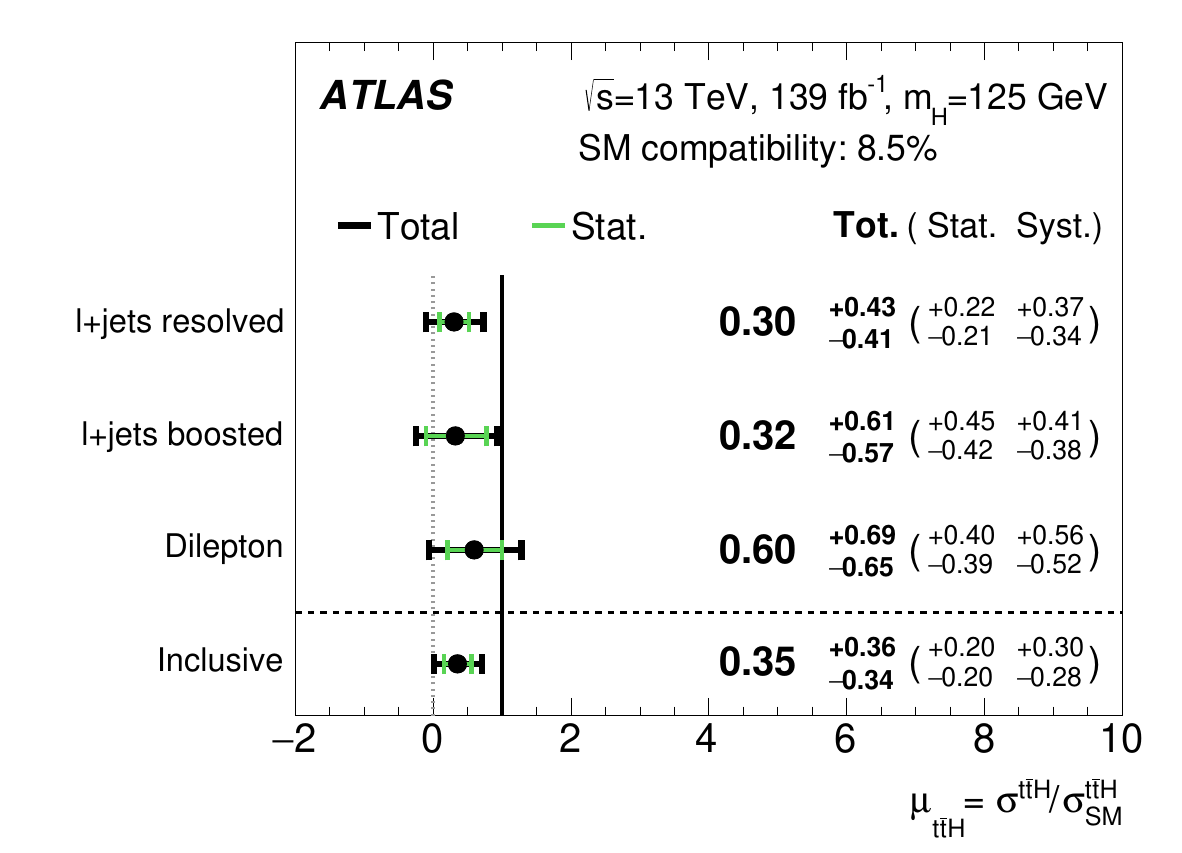}
	\includegraphics[width=7cm]{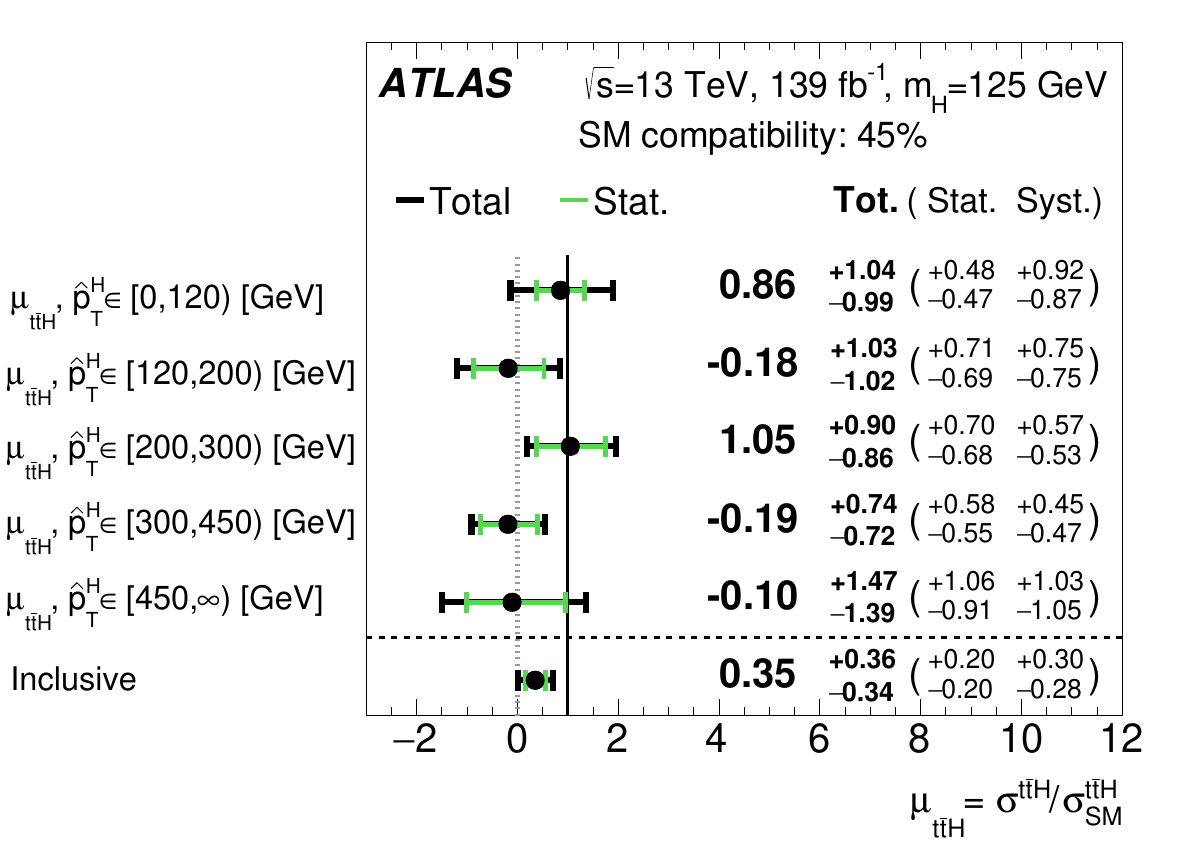}
	\includegraphics[width=7cm]{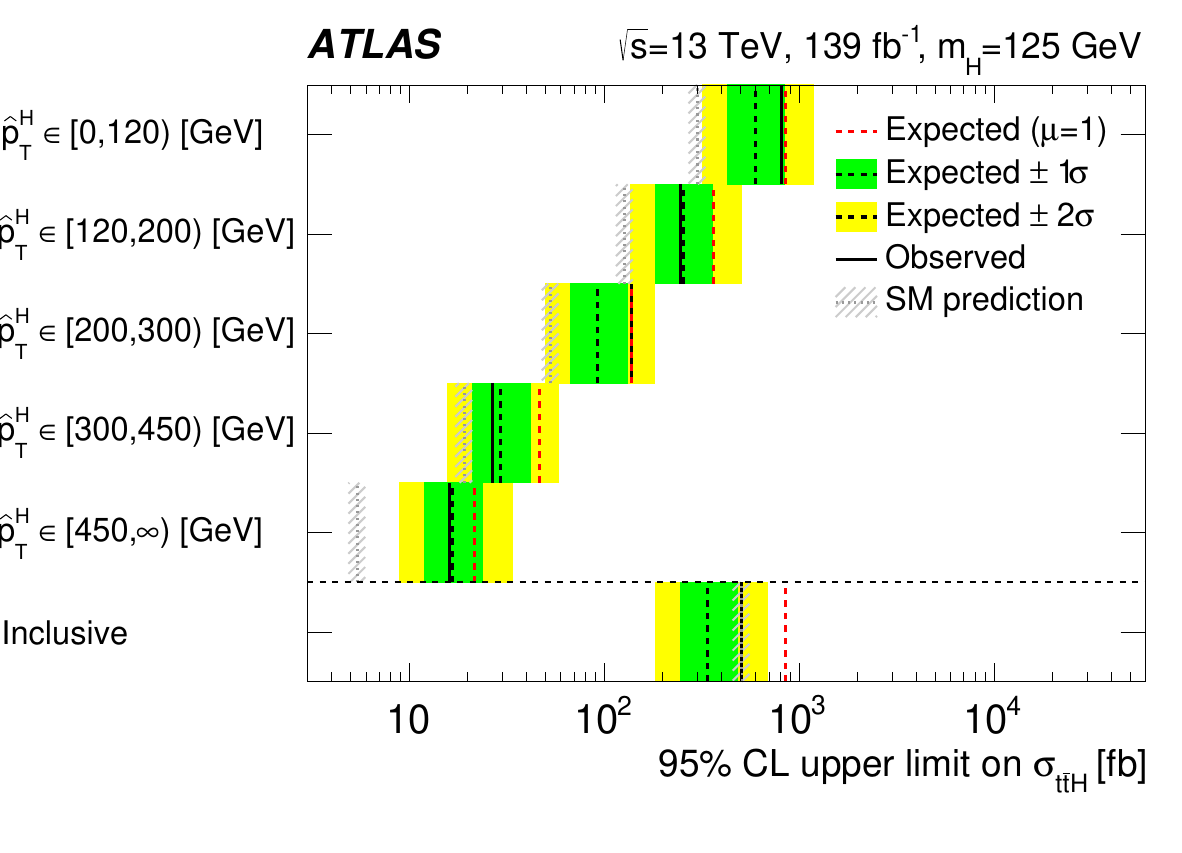}
	\caption{The best-fit values of the $t\bar{t}H$ signal strength in the individual channels as well as in the inclusive measurement (top-left). Signal-strength measurements (top-right) and 95\% CL upper limits in the individual STXS $p_\text{T}^H$ bins as well as the inclusive limit (bottom)~\cite{ttH}.}
	\label{ttH}
\end{figure}

The top-quark Yukawa coupling $(y_t)$ is the strongest coupling in the Standard Model. It is measured through the direct interaction of the top quark with the Higgs boson at the tree level. The analysis~\cite{ttH} is conducted in both the resolved and boosted regimes, with a distinction made between single-lepton and di-lepton regions. An excess of events beyond the expected Standard Model background is observed, corresponding to an observed (expected) significance of 1.0 (2.7) standard deviations. The measured signal strength is $\mu = 0.35 \pm 0.20 (stat.)^{+0.30}_{-0.28} (syst.)$. Figure~\ref{ttH} (top-left) shows the inclusive measurement of the signal strength as well as  in the individual channels. Additionally, the signal strength is measured in Higgs $p_\text{T}$ bins within the STXS framework, which is the first STXS measurement in the $t\bar{t}H$ channel and the STXS measurement in the $p_\text{T}^H >450$ GeV bin. Figure~\ref{ttH} (top-right) shows the values obtained. Furthermore, 95\% CL upper limits on the cross-section are measured in each STXS bin are shown in Figure~\ref{ttH} (bottom).

\section{$H\to \tau \tau$}
\begin{figure}[t]
	\centering
	\includegraphics[width=7.5cm]{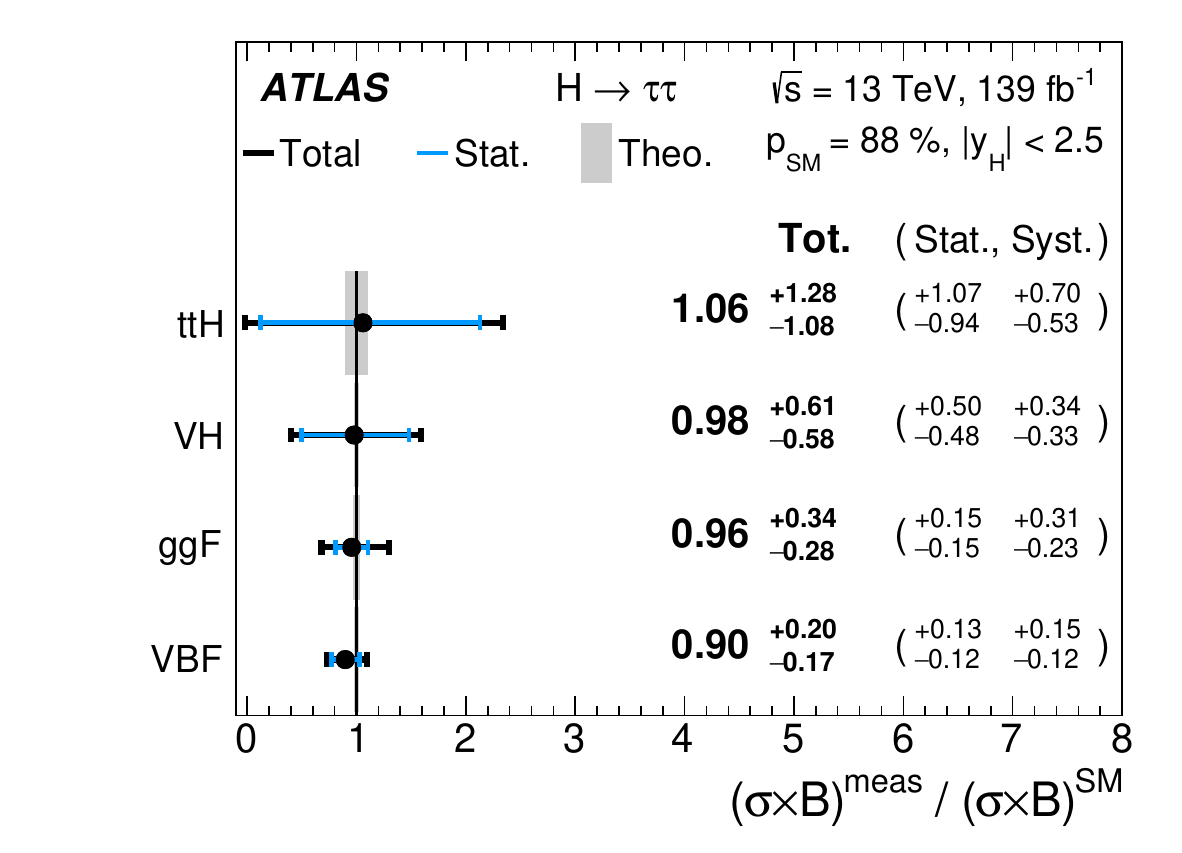}
	\includegraphics[width=7.5cm]{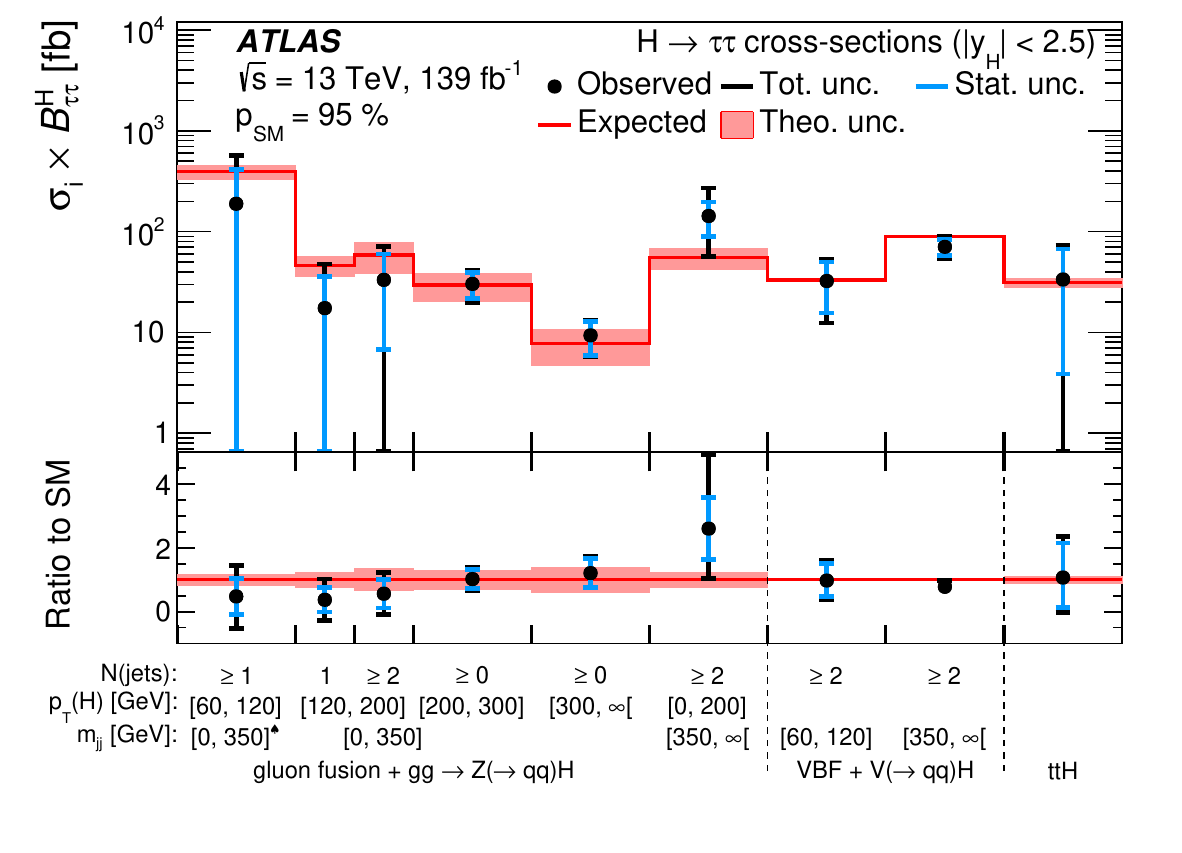}
	\caption{The measured values for $\sigma_H \times B(H\to \tau \tau)$ relative to the SM expectations, in the four dominant production modes (left), in the nine fiducial volumes defined in the STXS measurement (right)~\cite{htt}.}
	\label{fig:htt}
\end{figure}

 The cross section of the Higgs boson production is measured in the $H\to \tau \tau$ decay channel, targeting the four main production modes of the Higgs boson~\cite{htt}. The analysis includes three different channels: $\tau_{had} \tau_{had}$, $\tau_{lep} \tau_{had}$ and $\tau_e \tau_{\mu}$. The dominant backgrounds considered are $Z\to \tau \tau$, misidentified $\tau$, and top-quark production. The background from $Z\to \tau \tau$ is estimated using Monte Carlo simulation and dedicated control regions, while the contribution from misidentified $\tau$ is estimated using a data-driven approach. The total measured cross section for $H\to \tau \tau$ is found to be $2.94 \pm 0.21$ $(stat)^{+0.37}_{-0.32}$ ($syst$) pb, which is in agreement with the Standard Model prediction of $3.17 \pm 0.09$ pb. The $H\to \tau \tau$ cross sections are also measured as a function of $p_\text{T}^H$, $N_\mathrm{jets}$ $(p_\text{T} > 30$ GeV$)$ and $m_{jj}$ in a reduced set of the bins of the stage 1.2 of the STXS framework. The results, illustrated in Figure~\ref{fig:htt} (right). These measurements represent the most comprehensive investigation of Higgs boson production in the $\tau\tau$ decay channel until now. It also provides valuable information on the Higgs boson interactions with $\tau$ leptons and confirms the consistency between the observed data and the predictions of the Standard Model.

\section{Higgs-charm coupling}

\begin{figure}[t]
	\centering
	\includegraphics[width=7.5cm]{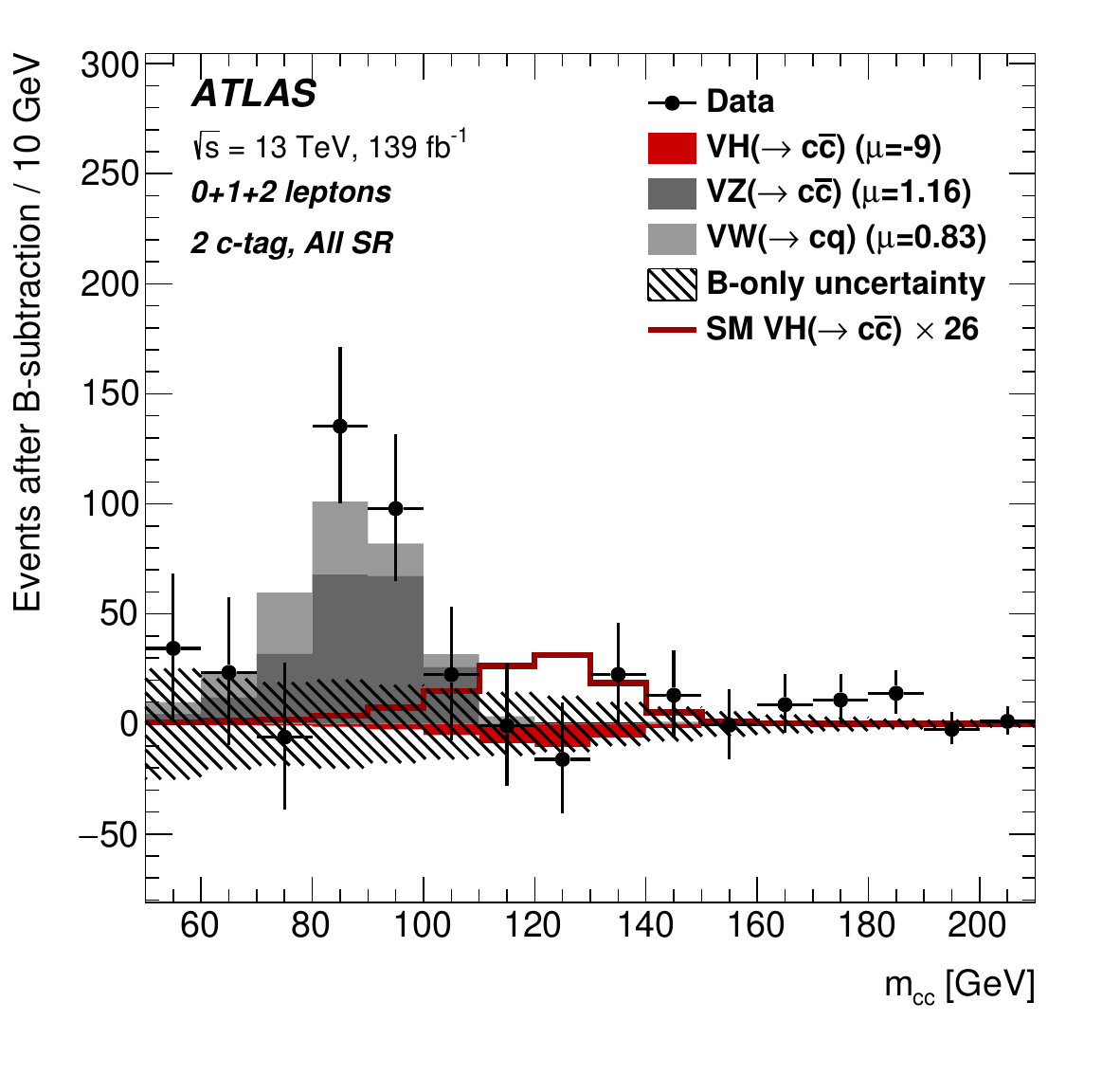}
	\includegraphics[width=7.5cm]{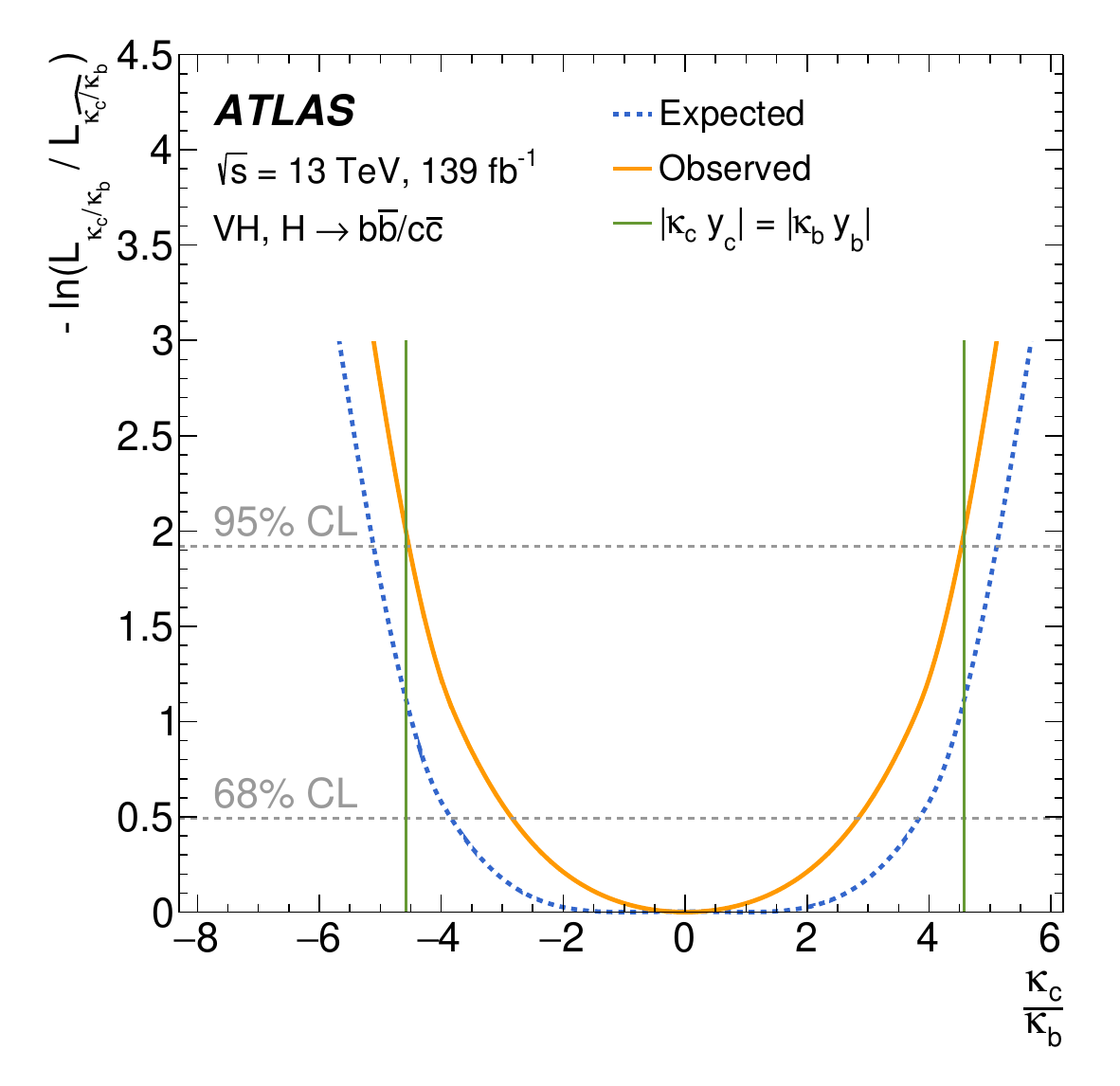}
	\caption{Signal strength ($\mu$) fit to $m_{c\bar{c}}$ in the search for $VH$, $H\to c\bar{c}$ events (left). Constraints on coupling modifier ratio $|\kappa_c/\kappa_b|$ obtained in a combination of $VH$, $H\to c\bar{c}$ and $VH$, $H\to c\bar{c}$ analyses (right)~\cite{hcc}.}
	\label{fig:hcc}
\end{figure}

\begin{figure}[h!]
	\centering
	\includegraphics[width=7.5cm]{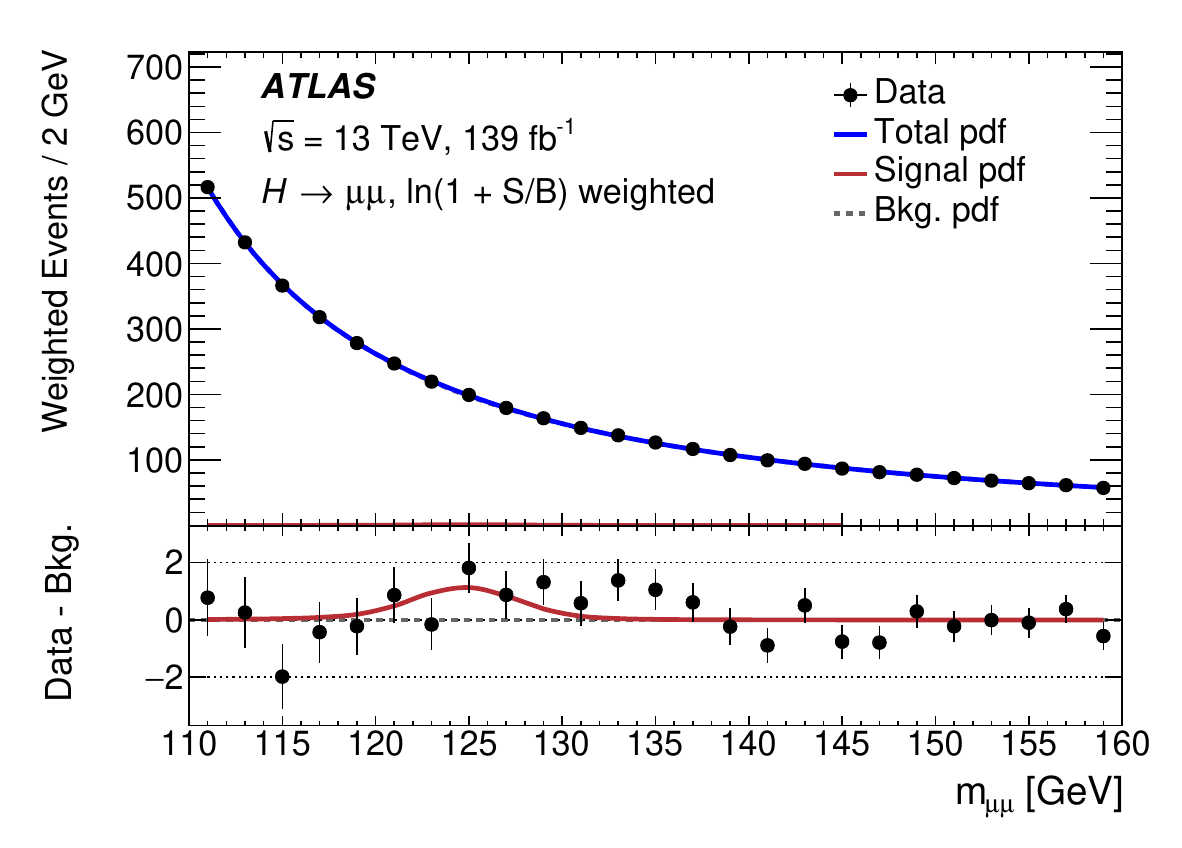}
	\includegraphics[width=7.5cm]{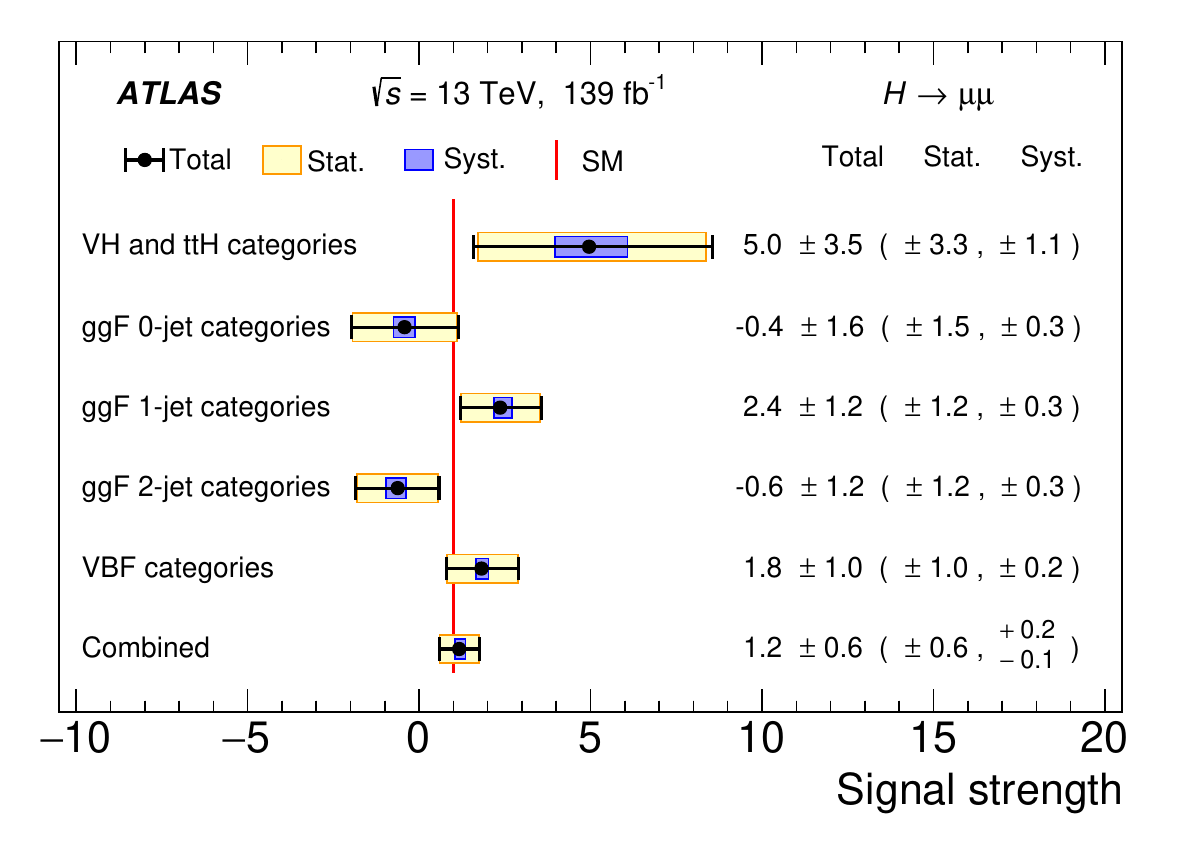}
	\caption{The spectrum of dimuon invariant mass in all the analysis categories observed in data (left). The best-ﬁt values of the signal strength parameters for the combination and ﬁve major groups of categories (right)~\cite{hmumu}.}
	\label{fig:hmumu}
\end{figure}
The search for $H\to c\bar{c}$~\cite{hcc} decays is carried out in the $VH$ production mode, which is considered to be the optimal production mode for studying $H\to b\bar{b}$ decays due to efficient triggering and background rejection facilitated by the leptonic decay of the weak gauge boson. However, the $H\to c\bar{c}$ analysis encounters challenges including a lower branching ratio of BR($H\to c\bar{c}$) = 2.89\%~\cite{corss-section}, approximately 20 times smaller than BR($H\to b\bar{b}$), and the need for charm tagging, which involves identifying charm-hadron decays with less distinct signatures compared to $b$-hadron decays. To address the difficulty of separating charm-jets from $b$-jets, a dedicated charm tagging algorithm is employed, which requires jets to have high c-tagging scores while vetoing jets with high $b$-tagging scores. This veto allows the combination of orthogonal signal events in the $H\to c\bar{c}$ and $H\to b\bar{b}$ analyses. The $VH$, $H\to c\bar{c}$ analysis extracts the signal and backgrounds through a simultaneous fit to the invariant mass of the $c$-tagged jets ($m_{c\bar{c}}$), resulting in the extraction of three signal strengths ($\mu$): $\mu_{VH(c\bar{c})}$, $\mu_{VW(cq)}$, and $\mu_{VZ(c\bar{c})}$. The best-fit distributions (Figure \ref{fig:hcc}: left) are compared to the data, and the solid line corresponds to the 95\% CL upper limit, $\mu_{VH(c\bar{c})}$ = 26.

A combination of $H\to c\bar{c}$ and $H\to b\bar{b}$ analyses is conducted to determine the constraints on the ratio $|\kappa_c/\kappa_b|$, where the coupling strength modifiers ($\kappa$) are defined as the ratios of the measured Yukawa couplings ($y$) to their Standard Model (SM) values: $\kappa_c = y_c/y_c^{SM}$ and $\kappa_b = y_b/y_b^{SM}$. The observed bound, shown in Figure \ref{fig:hcc} (right), yields $|\kappa_c/\kappa_b|$ $<$ 4.5 at 95\% CL. This bound is slightly smaller than the ratio $m_b/m_c$ indicated by vertical lines, indicating that the Higgs-charm coupling is weaker than the Higgs-bottom coupling. Notably, this result does not rely on any assumptions about the width of the Higgs boson, as the ratio approach allows the cancellation of the effects of Higgs total width on the coupling modifier determination.

\section{The elusive Higgs boson decay to muon pairs}
The decay process $H\to \mu \mu $~\cite{hmumu} offers a unique opportunity for investigating the coupling between the Higgs boson and second-generation fermions. The Standard Model predicts a branching ratio (BR) of $(2.17\pm0.04)\times10^{-4}$ for this decay mode. The dominant background for this process is the Drell-Yan process, with a typical signal-to-background ratio (S/B) at the level of 0.1\%. To enhance sensitivity, the analysis is divided into 20 different categories. According to the production modes, the signal events are exclusively selected, prioritizing the production mechanisms with increasing rarity but higher sensitivity to the Higgs boson decay to $\mu \mu$.

This result represents an improvement of about a factor of 2.5 in expected sensitivity compared to the previous publication based on 36 fb$^{-1}$ of data. Approximately 25\% of this improvement is attributed to the utilization of more advanced analysis techniques. The analysis includes categories that target all major production modes of the Higgs boson. In Figure~\ref{fig:hmumu} (left) the events in all 20 categories are weighted by ln(1 + S/B), where S are the observed signal yields and B are the background yields derived from the ﬁt to data. The observed data is found to be compatible with the Standard Model, with an observed significance of 2.0$\sigma$ (expected 1.7$\sigma$) and a signal strength measurement of $\mu = 1.2 \pm 0.6$~(Figure \ref*{fig:hmumu} right). Moreover, the analysis sets an upper limit on the branching ratio of $H\to \mu \mu $, excluding values above $4.7 \times 10^{-4}$ at 95\% CL.

\section{Summary}
 The study of the Higgs boson coupling to fermions has proven to be a crucial test for the Standard Model. The comprehensive analyses conducted by the ATLAS experiment, utilizing the full Run 2 datasets, consistently demonstrate that the Higgs boson coupling to fermions aligns with the predictions of the Standard Model. For analyses with adequate statistics and significance, the Simplified Template Cross Section (STXS) measurements with multiple Parameter of Interest (POIs) have been performed, and the results have been interpreted within the framework of Effective Field Theory (EFT). Overall, these analyses provide strong evidence of the agreement between experimental observations and the theoretical predictions of the Standard Model.

\section*{Acknowledgments}
The author is supported by the National Science Foundation of China under contract No. 12075060.


\begin{thebibliography}{99}
	
	\bibitem{ATLAS}
     ATLAS Collaboration, JINST 3 S08003 (2008).
     
    \bibitem{STXS}
     LHC Higgs Cross Section Working Group, LHCHXSWG-2019-003. 
    
    \bibitem{EFT}
     I. Brivio and M. Trott, Phys. Rept. 793 (2019) 1-98.
     
    \bibitem{corss-section}
     LHC Higgs Cross Section Working Group, CERN-2017-002-M.
    
    \bibitem{higgs}
     ATLAS, CMS Collaboration, Phys. Rev. Lett. 114 (2015) 191803.
     
    \bibitem{hbb1}
     ATLAS Collaboration, Eur. Phys. J. C 81 (2021) 178.

	\bibitem{hbb2}
	 ATLAS Collaboration, Phys. Lett. B 816 (2021) 136204.

	\bibitem{hbb3}
	 ATLAS Collaboration, ATLAS-CONF-2021-051, https://cds.cern.ch/record/2782535
 
    \bibitem{ttH}
     ATLAS Collaboration, JHEP 06 (2022) 097.
     
    \bibitem{htt}
     ATLAS Collaboration, JHEP 08 (2022) 175.

    \bibitem{hcc}
     ATLAS Collaboration, Eur. Phys. J. C 82 (2022) 717.

    \bibitem{hmumu}
     ATLAS Collaboration, Phys. Lett. B 812 (2021) 135980.
	
\end{thebibliography}
\end{document}